\documentclass{mem}
\usepackage{natbib}\usepackage{txfonts}\usepackage{balance}
\usepackage{graphicx}
\usepackage{amssymb}
\usepackage[a4paper]{hyperref}
\idline{75}{282}
\begin{document}
\def\teff{$T\rm_{eff }$}
\def\kms{$\mathrm {km s}^{-1}$}

\title{
On the 3D Structure of the Magnetic Field in Regions of Emerging Flux
}

   \subtitle{}

\author{
A. \,Asensio Ramos\inst{1,2} 
\and J. \,Trujillo Bueno\inst{1,2,3}
          }

  \offprints{A. Asensio Ramos}

\institute{
Instituto de Astrof\'{\i}sica de Canarias, 38205, La Laguna, Tenerife, Spain
\and
Departamento de Astrof\'{\i}sica, Universidad de La Laguna, 38205, Tenerife, Spain
\and
Consejo Superior de Investigaciones Cient\'{\i}ficas, Spain\\
\email{aasensio@iac.es, jtb@iac.es}
}

\authorrunning{Asensio Ramos \& Trujillo Bueno}

\titlerunning{Magnetic Structure of Emerging Flux Regions}

\abstract{We explore the photospheric and chromospheric magnetic field in an emerging flux region. An image of the
equivalent width of the He \textsc{i} 10830 \AA\ red blended component shows the presence of filamentary structures
that might be interpreted as magnetic loops. We point out that the magnetic field strength in the chromosphere 
resembles a smoothed version of that found in the photosphere and that it is not correlated at all
with the above-mentioned equivalent width map. Lacking other diagnostics, this suggests that one cannot discard
the possibility that the chromospheric field we infer from the observations 
is tracing the lower chromosphere of the active region instead of tracing the magnetic field along
loops. If the He \textsc{i} line is formed within magnetic loops, we point out a potential problem 
that appears when interpreting observations using only one component along the line-of-sight.
}
\maketitle{}

\section{Chromospheric magnetic fields}
Measuring magnetic fields at chromospheric and coronal heights is notoriously difficult
\citep[e.g., the reviews by][]{asensio-harvey06,asensio-casini_landi07,asensio-lagg_asr_07,asensio-trujillo09}.
There are two fundamental reasons for this. First, the number of lines with diagnostic
potential is very scarce. Second, obtaining physical information from the observed spectral line polarization
is complex, because chromospheric lines are often dominated by scattering
and they are sensitive to magnetic fields not only through the Zeeman effect but also through
the modification of the atomic level polarization by the Hanle effect.

\begin{figure*}[]
\resizebox{\hsize}{!}{\includegraphics[width=\textwidth]{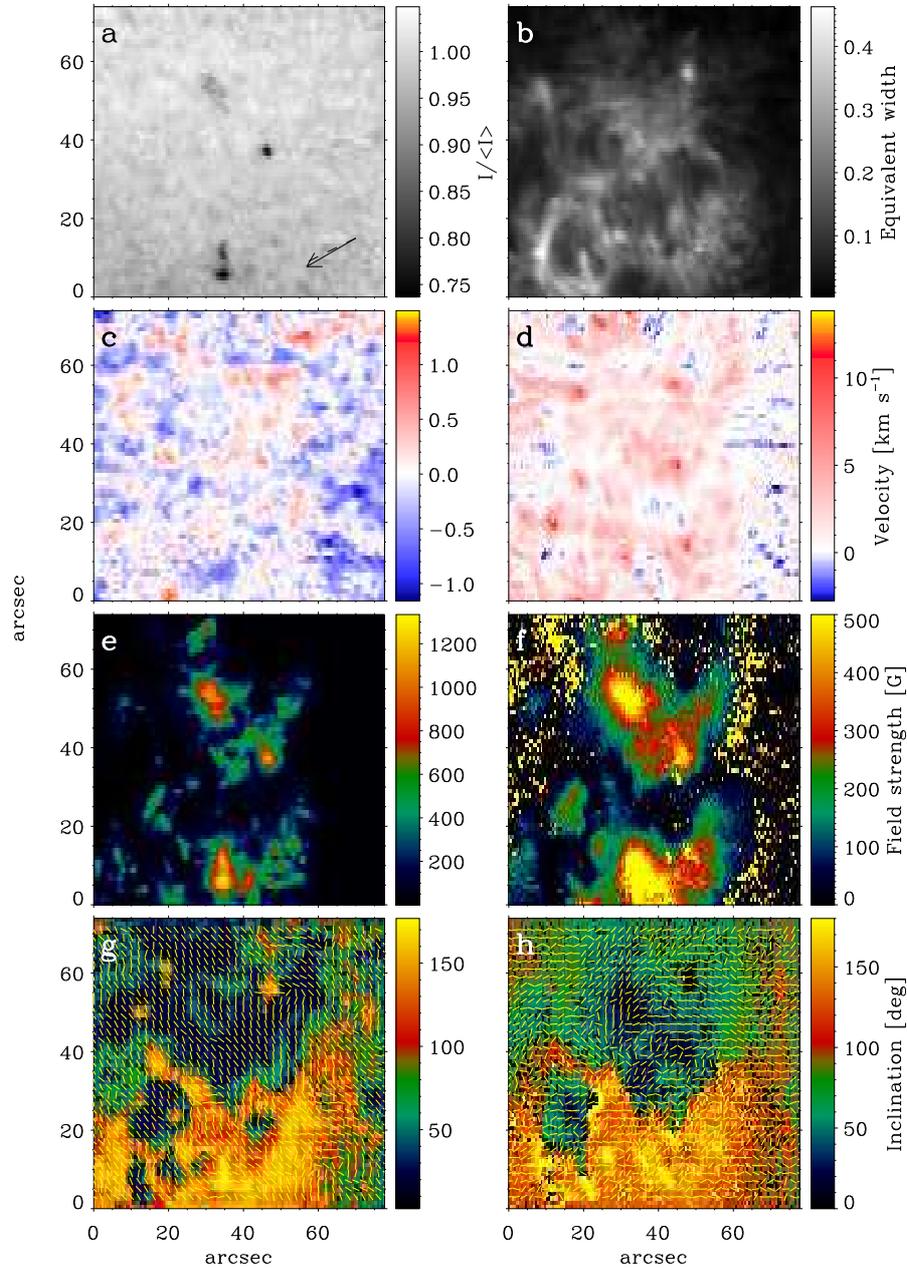}}
\caption{\footnotesize
Photospheric and chromospheric atmospheric parameters: (a) continuum image, (b) image of the equivalent width of the
He \textsc{i} red component, showing loop-like filamentary structures, (c) photospheric velocity
field, (d) chromospheric velocity field, showing strong downflows close to the footpoints of the loop-like
structures seen in the equivalent width image, (e,f) photospheric and chromospheric magnetic field strength, (g,h) 
magnetic field inclination in the background with an estimation of the azimuth (see the arrows). A fraction of the 
pixels have not been
used in the inversion of He \textsc{i} 10830 \AA\ triplet
because
the corresponding polarimetric signals are not above the prescribed threshold. (A color version of this
figure is available in the online version.)}
\label{asensio-fig:field_inferred}
\end{figure*}

In particular, the lines of neutral helium at 10830 \AA\ and 5876 \AA\ (D$_3$ multiplet) are of great interest
for empirical investigations of the 
dynamic and magnetic properties of plasma structures in the solar chromosphere
and corona, such as active 
regions \citep[e.g.,][]{asensio-harvey_hall71,asensio-ruedi96,asensio-Lagg04,asensio-centeno06}, filaments 
\citep[e.g.,][]{asensio-lin98,asensio-trujillo_nature02}, prominences \citep[e.g.,][]{asensio-landi_d3_82,asensio-querfeld85,asensio-bommier94,asensio-casini03,asensio-merenda06}
and spicules \citep[e.g.,][]{asensio-trujillo_merenda05,asensio-socas_elmore05,asensio-lopezariste_casini05,asensio-ramelli06,asensio-ramelli06_2,asensio-centeno09}.
The advantage of these spectral lines resides in the fact that they are almost absent in the quiet Sun and 
turn out to be relatively optically thin in the chromospheric and coronal structures where they originate. 

Recently, \cite{asensio-solanki_nature03} and \cite{asensio-Lagg04} reconstructed loop-like structures that arrive to the corona 
by assuming that the He \textsc{i} 10830 \AA\ triplet is formed within loops.
Since the analysis carried out by these authors depend on a prior assumption, \cite{asensio-judge09} has 
pointed out the possibility that the observations and the ensuing inferred magnetic field vector can be 
explained with simpler model assumptions: that the He \textsc{i} lines are formed in a corrugated
surface at chromospheric heights in the active region atmosphere. 

\begin{figure}[t!]
\resizebox{\hsize}{!}{\includegraphics[width=0.5\textwidth]{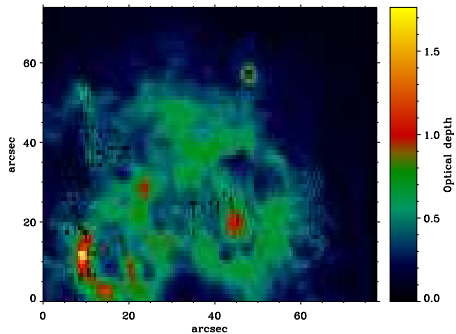}}
\caption{\footnotesize
Optical depth inferred at the center of the red component of the He \textsc{i} triplet. Note its similarity
with the map of equivalent width shown in panel b of Fig. \ref{asensio-fig:field_inferred}.}
\label{asensio-fig:filament_opticaldepth}
\end{figure}

Observations of an emerging flux region carried out during September 27, 2007 with the TIP-II polarimeter 
mounted on the German VTT have been analyzed using the 
HAZEL\footnote{\texttt{http://www.iac.es/project/magnetism}} 
inversion code \citep{asensio-asensio_trujillo_hazel08}
for the He \textsc{i} lines and the LILIA code \citep{asensio-socas_navarro01} for the photospheric Si \textsc{i} line.
The noise level is $\sim$6$\times$10$^{-4}$ in units of the continuum intensity, which
is only sufficient to detect the strongest Stokes $Q$ and $U$ signals in the red and blue components
of the He \textsc{i} 10830 \AA\ multiplet.
Our chromospheric inversions are carried out using a one-component constant-properties slab illuminated
from below by the photospheric radiation field, taking into account radiation transfer and magneto-optical
effects and calculating the effect of the magnetic field on the energy levels under the incomplete Paschen-Back
effect theory. The total computational time amounted to $\sim$72 hours in 4 processors, roughly 1--2 minutes per
pixel. 

Figure \ref{asensio-fig:field_inferred} presents a comparison between the inferred photospheric and chromospheric
physical parameters, together with the continuum image and the equivalent width (EW) map. The inversions show
several interesting features. First, strong downflows are detected at positions compatible with the endpoints
of the loop-like structures seen in the EW image, something already found by \cite{asensio-solanki_nature03} and \cite{asensio-Lagg04}.
Contrary to what they found, we do not clearly detect upflowing chromospheric material between the endpoints.
Second, the inferred photospheric and chromospheric field strength maps are very similar in appearance, although
in the chromospheric He \textsc{i} 10830 \AA\ triplet we detect field strengths 
a factor 2 smaller on average than in the photospheric Si \textsc{i} line. Apart from this, the chromospheric field strength 
distribution appears to be smoother than the photospheric one, in accordance with the higher formation heights. 
Third, it is interesting to point out that the filamentary structures seen 
in the EW map do not present a significantly different magnetic field strength. However, according to the results
presented in Fig. \ref{asensio-fig:filament_opticaldepth}, there is a very good correlation between the optical depth
inferred from the observations and the equivalent width. Consequently, the filamentary structure seen in the EW
image must be produced by
a density enhancement in a relatively uniform magnetic field.
Fourth, the inclination of the field is such that there is a relatively rapid transition from one polarity
to the other in the active region, both in the photosphere and in the chromosphere.

\section{Some Bayesian considerations}
The point raised by \cite{asensio-judge09} about the interpretation of the results of \cite{asensio-solanki_nature03} and 
\cite{asensio-Lagg04} is important and should be analyzed more deeply. Using standard Bayesian ideas \citep[see, e.g.,][]{asensio-jaynes03}, 
the problem is equivalent to that of comparing hypothesis $H_0$ (He \textsc{i} lines
are formed in a loop) with hypothesis $H_1$ (He \textsc{i} lines are formed in a horizontal slab) for the
explanation of a set of observations $D$. Model comparison should be carried out by calculating the ratio of
posteriors for each hypothesis, which simplifies, thanks to the Bayes theorem, to the product of the
ratio of evidences and the ratio of priors:
\begin{equation}
R = \frac{p(H_0|D)}{p(H_1|D)} = \frac{p(D|H_0)}{p(D|H_1)} \times \frac{p(H_1)}{p(H_0)},
\end{equation}
where
\begin{equation}
p(D|H) = \int p(D|\theta,H) p(\theta|M) \mathrm{d}\theta,
\end{equation}
with $\theta$ the set of parameters defining model $H$. Whether or not hypothesis $H_0$ is to be preferred with respect to hypothesis
$H_1$ can be established by the so-called ``Jeffreys' scale'' \citep[see, e.g.,][]{asensio-trotta08}.
We have weak evidence if $R \sim 3$, moderate evidence if $R \sim 12$ and strong evidence if $R \sim 150$, while
it remains inconclusive if $R \lesssim 3$. Both hypotheses present the very same number of parameters and,
in principle, both of them fit the data equally well, so that one could assume that both evidences
are relatively similar. This demonstrates that, deciding whether He \textsc{i} lines
are formed in a loop or in a horizontal slab is a matter of prior knowledge. It is justified to say
that both models are equally probable because $R \sim 1$. Obviously, a way to overcome this situation is, 
as suggested by \cite{asensio-judge09}, to augment the problem with new data $D$, so that the ratio of evidences
clearly favors one hypothesis. The reason is that, in such a case, one model would fit better
the new observables (e.g., stereoscopic observations).

\section{Additional complications and conclusions}
\begin{figure*}[t!]
\resizebox{\hsize}{!}{\includegraphics[width=0.8\textwidth]{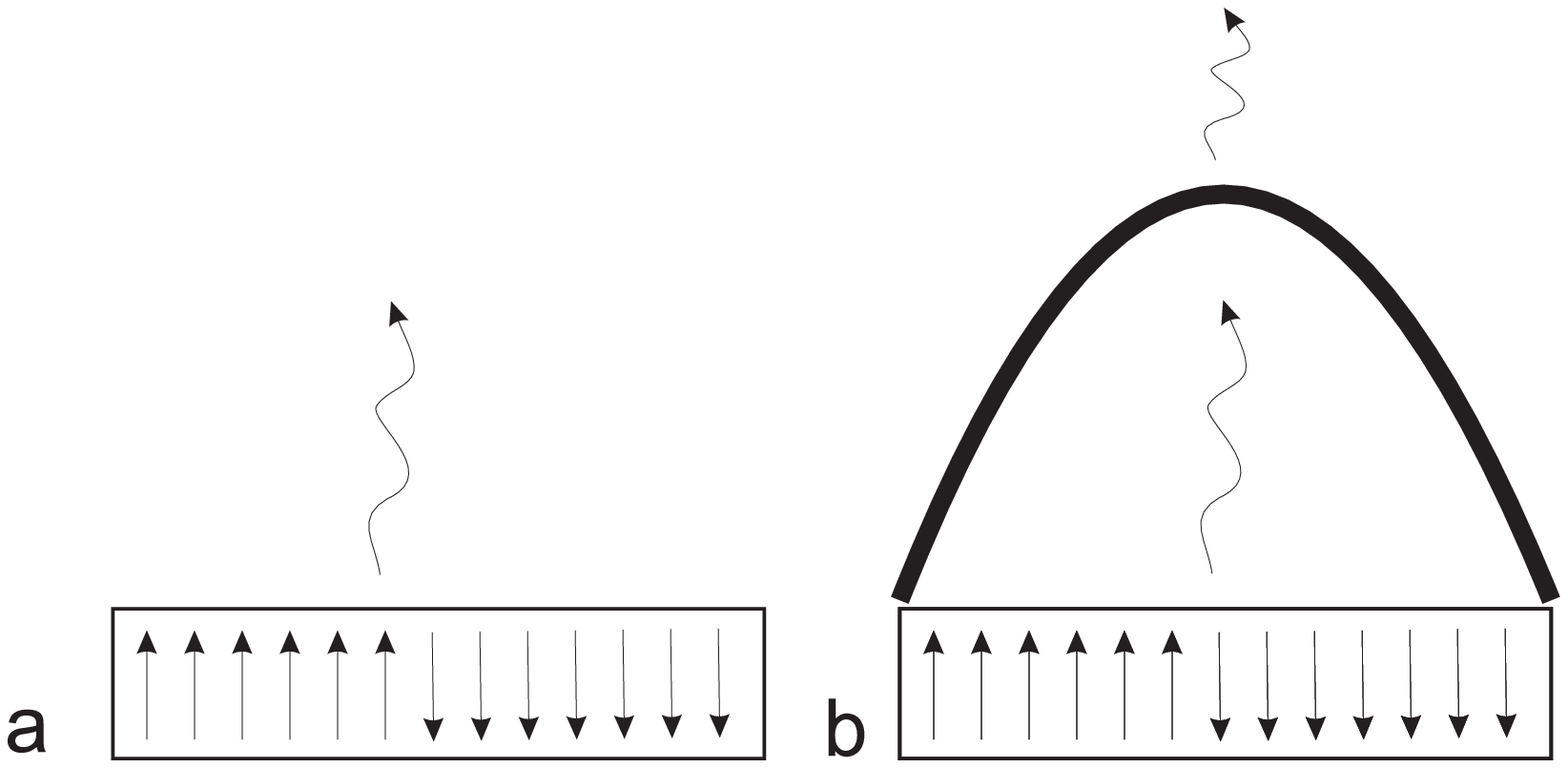}}
\resizebox{\hsize}{!}{\includegraphics[width=0.5\textwidth]{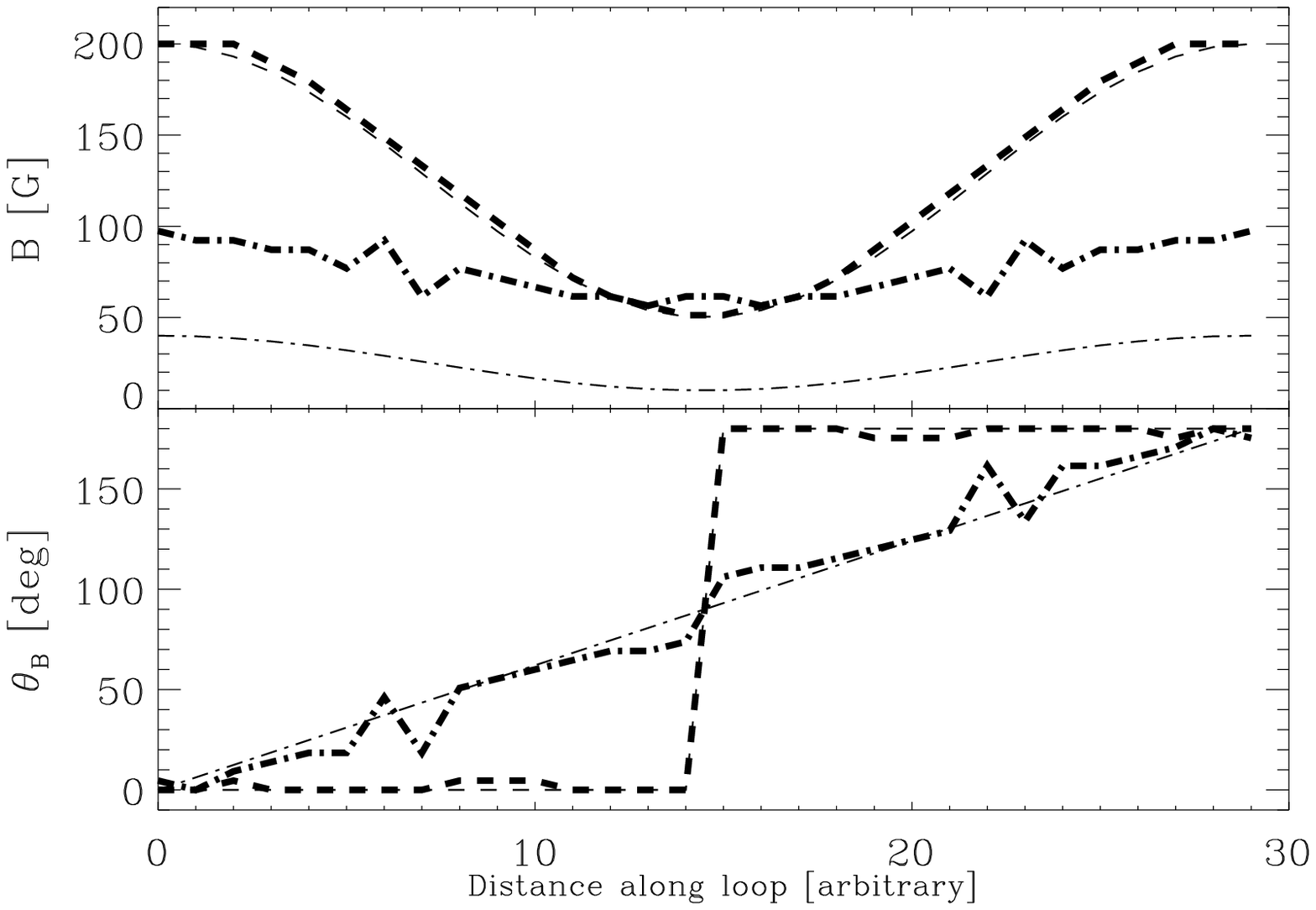}\includegraphics[width=0.5\textwidth]{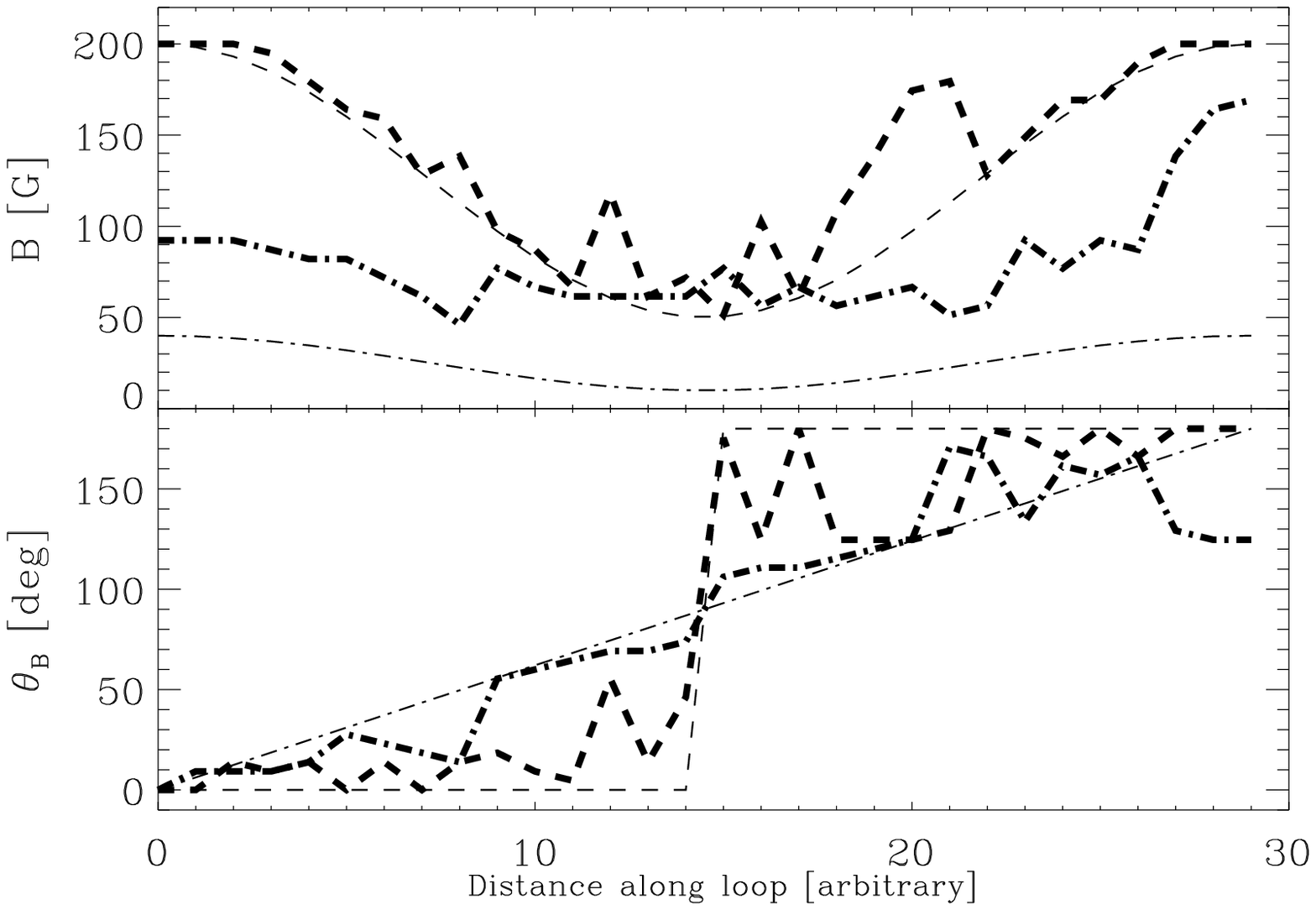}}
\caption{\footnotesize
Upper panels: sketch showing the two possible situations encountered when observing an emerging flux region
with an active chromosphere and a loop, both producing He \textsc{i} absorption. Lower panels: input
field strength and inclination, and the inferred ones using the inversion code HAZEL for the case without noise (lower left panel) and
with a noise similar to our observations (lower right panel). The thin dashed lines show the input values for the
chromospheric slab, while the thin dot-dashed present the input for the loop. The thick curves indicate
what one infers from the observations in cases (a) and (b), respectively.}
\label{asensio-fig:filament_inference}
\end{figure*}

As shown by \cite{asensio-solanki_nature03}, \cite{asensio-Lagg04} and in the present work, the chromosphere of the observed region where
weak He \textsc{i} absorption can be found is still magnetized. Consequently,
even if loop-like structures reaching coronal heights exist, the interpretation of the polarimetric signals in
terms of a slab of constant physical properties or in terms of a Milne-Eddington model 
is dubious and needs consideration. The reason is that, as shown in
the upper panel of Fig. \ref{asensio-fig:filament_inference}, we can encounter two situations: (a) one in which rays along the line-of-sight
do not cross any loop and we see directly the chromosphere of the active region and (b) another one in which rays along 
the line-of-sight cross a loop before escaping. 

We first focus on case (a) and analyze what happens as we move from one region with a given
polarity to the other one with the opposite polarity. We assume that the chromosphere can be modeled as 
a constant properties slab whose magnetic field is
either pointing upwards or downwards. The change in inclination is sharp at the 
central part of the slab and the field strength
changes smoothly from 200 G at the center of each magnetic region to 50 G in the middle where the polarity reversal takes place. The
input variation of the field strength and inclination with distance along the line joining the two polarities is shown in the lower
panels of Fig. \ref{asensio-fig:filament_inference} (see the thin dashed lines). Using HAZEL, we synthesize the Stokes profiles emerging
at each point assuming that the slab has an optical depth of $\Delta \tau=0.3$ measured at the center of the red blended component of the
He \textsc{i} 10830 \AA\ multiplet. The field strength and inclination of the field 
are inferred using HAZEL keeping fixed the rest of parameters to the
correct values. The blue solid curve in the lower left panel shows the inferred values in the case without noise, while the 
lower right panel presents the results when a noise level similar to our observations is added. We note that the 
inferred values are very close to the input ones, especially in the case without noise. In the noisy case, some
fluctuations induce that the best fit in some points is achieved with a slightly larger field and a slightly more
inclined field but the fundamental characteristics are recovered.

In case (b), the radiation escaping from the lower slab goes through a loop of optical depth $\Delta \tau=0.6$. The
magnetic field strength along the loop changes smoothly from 40 G close to the endpoints to 10 G in the central part.
The inclination of the field is assumed to vary linearly, being horizontal at the top of the loop. The thin dot-dashed curves
in the lower panels of Fig. \ref{asensio-fig:filament_inference} present such variations. Following the same strategy, we
apply HAZEL to synthesize the emergent intensity and polarization taking into account that the 
Stokes profiles entering the loop are those emerging from the lower slab. We apply HAZEL to infer the magnetic 
field vector from the synthetic observations and the results are shown in thick lines. 
For simplicity, we fix the total optical depth of the slab to $\Delta \tau=0.9$.
We have verified that this is the optical depth inferred from the synthetic observations if we leave this parameter free.
Since we force HAZEL to interpret with only one field vector the combination of Stokes profiles 
produced by two completely different field
distributions, the inferred magnetic field vector is somewhat between that in the lower
slab and that in the loop, as indicated by the red curves. Note that this happens even in the case without 
noise. As a consequence, if a loop producing absorption in He \textsc{i} is placed above an active region
whose chromosphere also produces a significant absorption, one should be careful with the interpretation
given to the inferred magnetic field vector.
 
\begin{acknowledgements}
Financial support by the Spanish Ministry of Science and Innovation through project AYA2007-63881 is gratefully acknowledged.
\end{acknowledgements}


\end{document}